\documentclass[twocolumn,letterpaper,aps,prl,superscriptaddress,showpacs,nofootinbib,floatfix]{revtex4}

\usepackage{graphicx}
\usepackage{slashbox}
\usepackage{brackets}
\usepackage{psfrag}
\usepackage{multirow}
\usepackage{tabularx}
\usepackage{comment}
\begin{document}

\ProvideTextCommandDefault{\textonehalf}{${}^1\!/\!{}_2\ $}

\title{Measurement of the Doubly-Polarized $\vec{{^3}He}(\vec{\gamma},n)pp$ Reaction at 16.5 MeV and Its Implications for the GDH Sum Rule}

\author{G.~Laskaris}\email[Electronic address: ]{laskaris@stanford.edu}
\altaffiliation[Currently at ]{Stanford University, Stanford, CA 94305, USA}
\affiliation{Triangle Universities Nuclear Laboratory, Durham, North Carolina 27708, USA}
\affiliation{Department of Physics, Duke University, Durham, North Carolina 27708, USA}
\author{X.~Yan}
\affiliation{Triangle Universities Nuclear Laboratory, Durham, North Carolina 27708, USA}
\affiliation{Department of Physics, Duke University, Durham, North Carolina 27708, USA}
\author{J.~M.~Mueller}
\altaffiliation[Currently at ]{NCSU, Raleigh, NC 27695, USA}
\affiliation{Triangle Universities Nuclear Laboratory, Durham, North Carolina 27708, USA}
\affiliation{Department of Physics, Duke University, Durham, North Carolina 27708, USA}
\author{W.~R.~Zimmerman}
\affiliation{Triangle Universities Nuclear Laboratory, Durham, North Carolina 27708, USA}
\affiliation{Department of Physics, Duke University, Durham, North Carolina 27708, USA}
\author{W.~Xiong}
\affiliation{Triangle Universities Nuclear Laboratory, Durham, North Carolina 27708, USA}
\affiliation{Department of Physics, Duke University, Durham, North Carolina 27708, USA}
\author{M.~W.~Ahmed}
\affiliation{Triangle Universities Nuclear Laboratory, Durham, North Carolina 27708, USA}
\affiliation{Department of Physics, Duke University, Durham, North Carolina 27708, USA}
\affiliation{Department of Mathematics and Physics, North Carolina Central University, Durham, North Carolina 27707, USA}
\author{T.~Averett}
\affiliation{College of William and Mary, Williamsburg, Virginia 23187, USA}
\author{P.-H.~Chu}
\altaffiliation[Currently at ]{LANL, Los Alamos, NM 87544, USA}
\affiliation{Triangle Universities Nuclear Laboratory, Durham, North Carolina 27708, USA}
\affiliation{Department of Physics, Duke University, Durham, North Carolina 27708, USA}
\author{A.~Deltuva}
\affiliation{Institute of Theoretical Physics and Astronomy, Vilnius University,  LT-01108 Vilnius, Lithuania}
\author{C.~Flower}
\affiliation{Triangle Universities Nuclear Laboratory, Durham, North Carolina 27708, USA}
\affiliation{Department of Physics, Duke University, Durham, North Carolina 27708, USA}
\author{A.~C.~Fonseca}
\affiliation{Centro de F\'{i}sica Nuclear da Universidade de Lisboa, P-1649-003 Lisboa, Portugal}
\author{H.~Gao}
\affiliation{Triangle Universities Nuclear Laboratory, Durham, North Carolina 27708, USA}
\affiliation{Department of Physics, Duke University, Durham, North Carolina 27708, USA}
\author{J.~Golak}
\affiliation{M. Smoluchowski Institute of Physics, Jagiellonian University, PL-30348 Krak\'{o}w, Poland}
\author{J.~N.~Heideman}
\affiliation{Triangle Universities Nuclear Laboratory, Durham, North Carolina 27708, USA}
\affiliation{Department of Physics and Astronomy, University of North Carolina at Chapel Hill, Chapel Hill, North Carolina 27599, USA}
\author{ H.~J.~Karwowski}
\affiliation{Triangle Universities Nuclear Laboratory, Durham, North Carolina 27708, USA}
\affiliation{Department of Physics and Astronomy, University of North Carolina at Chapel Hill, Chapel Hill, North Carolina 27599, USA}
\author{M.~Meziane}
\affiliation{Triangle Universities Nuclear Laboratory, Durham, North Carolina 27708, USA}
\affiliation{Department of Physics, Duke University, Durham, North Carolina 27708, USA}
\author{P.~U.~Sauer}
\affiliation{Institut f\"ur Theoretische Physik, Leibniz Universit\"at Hannover, D-30167 Hannover, Germany}
\author{R.~Skibi\'nski}
\affiliation{M. Smoluchowski Institute of Physics, Jagiellonian University, PL-30348 Krak\'{o}w, Poland}
\author{I.~I.~Strakovsky}
\affiliation{Department of Physics, The George Washington University, Washington DC 20052, USA}
\author{H.~R.~Weller}
\affiliation{Triangle Universities Nuclear Laboratory, Durham, North Carolina 27708, USA}
\affiliation{Department of Physics, Duke University, Durham, North Carolina 27708, USA}
\author{H.~Wita{\l}a}
\affiliation{M. Smoluchowski Institute of Physics, Jagiellonian University, PL-30348 Krak\'{o}w, Poland}
\author{Y.~K.~Wu}
\affiliation{Triangle Universities Nuclear Laboratory, Durham, North Carolina 27708, USA}
\affiliation{Department of Physics, Duke University, Durham, North Carolina 27708, USA}

\date{\today}

\begin{abstract}
We report new measurements of the doubly-polarized photodisintegration of $^3$He at an incident photon energy of 16.5 MeV, carried out at the High Intensity $\gamma$-ray Source (HI$\gamma$S) facility located at Triangle Universities Nuclear Laboratory (TUNL). The spin-dependent double-differential cross sections and the contribution from the three--body channel to the Gerasimov-Drell-Hearn (GDH) integrand were extracted and compared with the state-of-the-art three--body calculations. The calculations, which include the Coulomb interaction and are in good agreement with the results of previous measurements at 12.8 and 14.7 MeV, can no longer describe the cross section results at 16.5 MeV. The GDH integrand was found to be about one standard deviation larger than the maximum value predicted by the theories. 
\end{abstract}


\maketitle

An important window for the study of QCD is through the investigation of the structure and particularly the spin structure of the nucleon and few-body nuclei. Therefore sum rules involving the spin structure of the nucleon or nuclei are nowadays at the forefront of intensive experimental and theoretical efforts. Among spin sum rules, the GDH sum rule~\cite{Drell} is particularly interesting. This sum rule relates the energy-weighted difference of the spin-dependent total photo-absorption cross sections $\sigma^P$ ($\sigma^A$) for target spin and beam helicity parallel (antiparallel) to static properties of the target nucleon/nucleus, i.e. the anomalous magnetic moment and the mass, as follows:
\begin{equation}
I^{GDH} = \int_{\nu _{thr}}^{\infty}(\sigma^P- \sigma^A)
{\frac{d\nu}{\nu}} = \frac{4\pi^{2}e^{2}}{M^{2}}\kappa^{2} I,
\label{Igdhr}
\end{equation}
where $\nu$ is the photon energy, $\nu_ {thr}$ is the pion production/photodisintegration threshold on the nucleon/nucleus, $\kappa$ is the anomalous magnetic moment, $M$ is the mass and $I$ is the spin of the nucleon/nucleus. There have been worldwide efforts in testing the GDH sum rule on proton and neutron~\cite{GDH-p,GDH-n}. More recently, experimental investigations of the GDH sum rule on nuclei such as the deuteron~\cite{Slifer2008,Ahmed,Ahrens2009} and $^3$He~\cite{GeorgePRL,laskaristhesis,Mainz3He,Mainz3He2} have begun.

The determination of the GDH sum rule on $^3$He at the energy region between the two-body photodisintegration ($\sim$5.5 MeV) and the pion production threshold ($\sim$140 MeV) is particularly interesting for a number of reasons. This energy region has an important contribution to the overall sum rule~\cite{gaoproc,laskaristhesis} and it is a region where one can test state-of-the-art three--body calculations. The experimental determination of the GDH integral on $^3$He can also test to what extent a polarized $^3$He target is an effective polarized neutron target. A polarized $^3$He target is commonly used as a polarized neutron target to extract the electromagnetic form factors~\cite{Gao,Xu,Riordan} and the spin structure functions~\cite{Anthony} of the neutron since the nuclear spin of $^3$He is carried mostly by the unpaired neutron. To acquire information about the neutron using a polarized $^3$He target, nuclear corrections relying on three--body calculations need to be used, but first they must be validated by experiments.

The GDH integral below pion threshold can be estimated based on three--body calculations which are performed mainly through the machinery of Faddeev~\cite{fad} and Alt-Grassberger-Sandhas equations (AGS)~\cite{Alt} and have been carried out for both two-body and three--body photodisintegration of $^3$He with double polarizations. These calculations~\cite{Deltuva,Skibinski} use a variety of nucleon-nucleon (NN) potentials like Argonne V18 (AV18)~\cite{wiringa} or CD Bonn~\cite{Machleidt19871,machle} and three-nucleon forces (3NFs) like Urbana IX (UIX)~\cite{Carl} or CD Bonn + $\Delta$~\cite{Deltuva}, with the latter yielding an effective 3NF through the $\Delta$-isobar excitation. The plateau value that both sets of calculations~\cite{Deltuva,Skibinski} predict for the GDH integral of $^3$He below pion threshold is $\sim$140 $\mu$b~\cite{laskaristhesis}. This part equals the sum of the contributions from the three--body $\sim$170~$\mu$b ($\sim$130~$\mu$b) and the two-body $\sim$-30 $\mu$b ($\sim$10 $\mu$b) components based on the calculations of Ref.~\cite{Deltuva} (Ref.~\cite{Skibinski}). 

The first experiment~\cite{GeorgePRL,laskaristhesis} on the three--body photodisintegration of $^3$He using a longitudinally polarized $^3$He target and a circularly polarized $\gamma$-ray beam took place at the HI$\gamma$S facility~\cite{higsreview} of TUNL at the incident photon energies of 12.8 and 14.7 MeV. The AGS calculations~\cite{Deltuva} including single-baryon and meson-exchange electromagnetic currents (MEC), relativistic single-nucleon charge corrections (RC)~\cite{Deltuva} and the proton-proton Coulomb force using the method of screening and renormalization~\cite{Deltuva}, provided a good description of the results.

\begin{figure*}[!ht]
  \centering
    \includegraphics[width=1.0\textwidth]{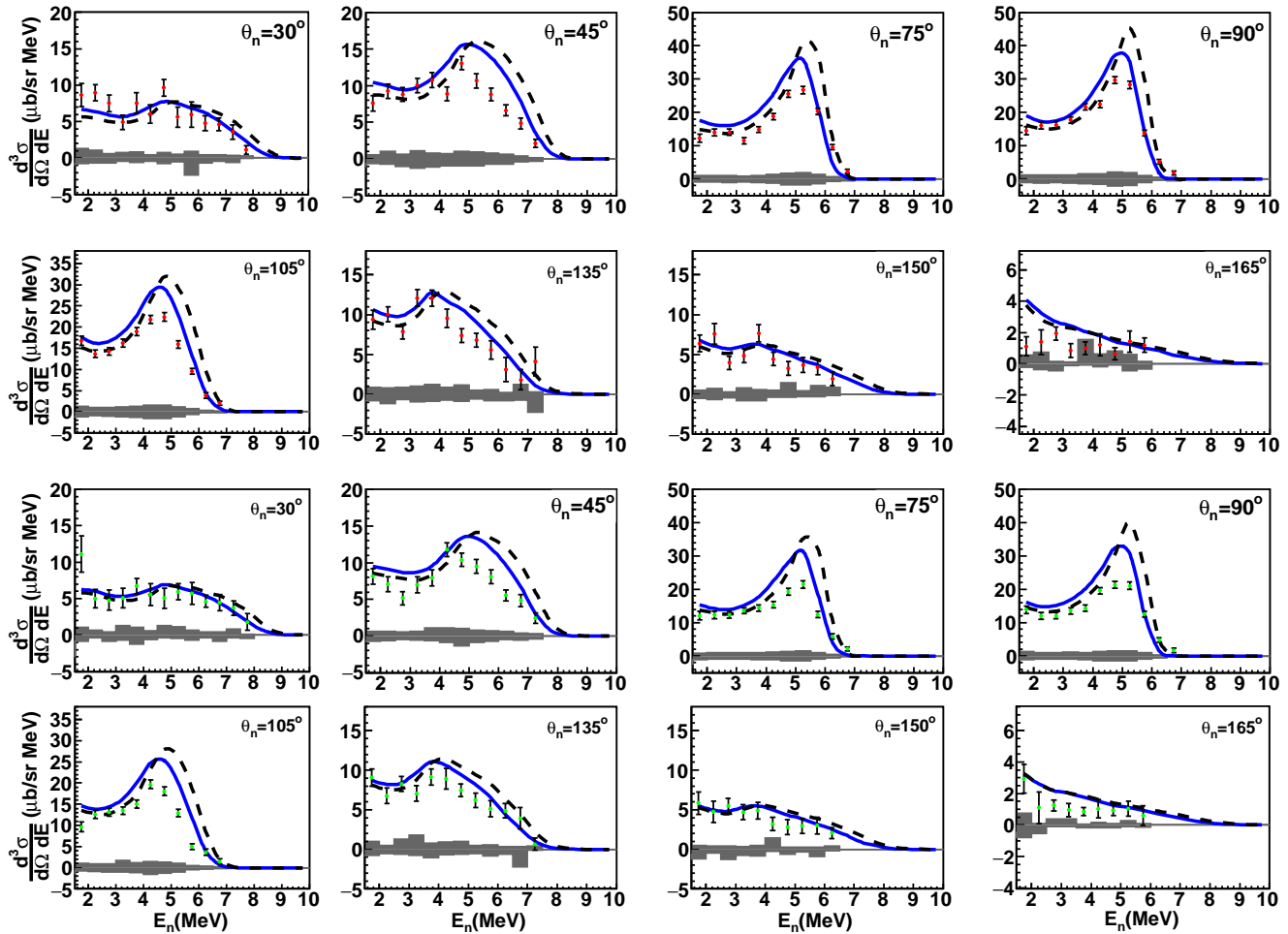}
    \caption{(Color online) Spin-dependent double-differential cross sections for the extended target for both parallel (two top rows) and antiparallel (two bottom rows) spin states as a function of the neutron energy, E$_n$, at $\nu$=16.5 MeV. The solid-blue curve is the calculation based on Ref.~\cite{Deltuva} including CD Bonn + $\Delta$-isobar + RC + MEC + Coulomb force while the dashed-black curve is from Ref.~\cite{Skibinski} including AV18 + UIX + MEC. The neutron energy bin width is 0.5 MeV. The band shows the combined systematic uncertainties.}
   \label{fig:165}
\end{figure*}

To investigate further whether such an agreement continues as one goes to higher energy and resolve the discrepancy between the past unpolarized cross section measurements above 15 MeV as it was shown in Ref.~\cite{GeorgePRL}, a new measurement of $\vec{^3He}(\vec{\gamma},n)pp$ was performed at the incident photon energy of 16.5 MeV and it is reported in this Letter. As in the previous experiment~\cite{GeorgePRL,laskaristhesis}, a nearly mono-energetic, $\sim$100\% circularly-polarized pulsed $\gamma$-ray beam was used. The beam was collimated using a 12 mm diameter collimator resulting in on-target intensities of (7.3-9.5)$\times10^{7}\gamma/s$ and an energy spread of $\Delta \nu/\nu\leq$5.0\%. A 10.6 cm long C$_6$D$_6$ cell was placed in the beam downstream of the target and two BC501A liquid scintillator neutron detectors were mounted at a scattering angle of $90\,^{\circ}$ degrees to detect the neutrons from deuteron photodisintegration. The on-target intensity of the beam was determined using the well-known d($\gamma$,n)p cross section~\cite{d2o}.

Upstream of the flux monitor, the polarized $\gamma$-beam was incident on a polarized $^3$He cell. The $^3$He and N$_2$ reference cells used for background subtraction were the same as in the previous experiment~\cite{GeorgePRL,laskaristhesis}. Details concerning their technical characteristics and the spin exchange optical pumping technique used to polarize the $^3$He target can be found in Refs.~\cite{Happer,Kramer,Ye,GeorgePRL,laskaristhesis}. The spin of the  $^3$He target was flipped every 15 m in order to extract the spin dependent cross sections and the GDH integrand, $(\sigma^P-\sigma^A)/\nu$. The polarization was measured using the nuclear magnetic resonance-adiabatic fast passage~\cite{Lorenzon} technique calibrated by electron paramagnetic resonance~\cite{epr4}. The latter can measure the absolute $^3$He target polarization, $P_{t}$ which was found to be between 33\% and 37\%.

An array of sixteen liquid scintillator BC-501A counters was used to detect the neutrons from the $\vec{{^3}He}(\vec{\gamma},n)pp$ reaction. The detectors were placed at the horizontal plane every $15\,^{\circ}$, symmetrically on each side of the beam, at laboratory scattering angles from $30\,^{\circ}$ to $165\,^{\circ}$. No detectors were placed at the laboratory angles $60\,^{\circ}$ and $120\,^{\circ}$ due to the proximity to a pair of Helmholtz coils which provided the holding field for the polarized $^3$He target.

Three quantities were recorded for each event: the pulse height (PH), the time-of-flight (TOF) and the pulse shape discrimination (PSD) signals. Initially, a PH cut was applied at 0.162 MeVee to set the detector efficiency. The correlations between the PSD, PH and TOF were utilized to extract the neutron events and to remove the $\gamma$-ray events and two-dimensional cuts were applied on these histograms. The same cuts were used for the data taken with the N$_2$ reference cell to subtract the background contributions. The outgoing neutron energy was determined using the measured TOF of the neutrons assuming they were emitted from the center of the $^3$He target cell. The neutron detection efficiency varied rapidly as a function of neutron energy below 2.0 MeV. Therefore, we report cross sections only for neutrons with kinetic energies above 2.0 MeV. More details about this analysis can be found in Refs.~\cite{GeorgePRL,laskaristhesis}.

The measured neutron background-subtracted yields ($^3$He neutron events/N$_\gamma$) at the $i^{th}$ energy bin for target spin parallel/anti-parallel to the helicity of the beam were calculated as $Y_{i,m}^{P/A}=Y_{i}^{P/A,^3He}-Y_{i}^{N_2}$, where $Y_{i}^{P/A,^3He}$ and $Y_{i}^{N_2}$ were the yields of reactions on $^3$He and N$_2$ cells. Their linear combination led to the yields for parallel and antiparallel spin-helicity states $Y_{i}^{P/A}=\frac{1}{2}(Y_{i,m}^{P}(1\pm \frac{1}{P_t P_b})+Y_{i,m}^{A}(1\mp \frac{1}{P_t P_b}))$, where $P_b$ is the beam polarization. The double-differential cross sections were defined as
\begin{equation}
\frac{d^{3}\sigma^{P/A}}{d\Omega dE_{n}}=\frac{Y_{i}^{P/A}}{\Delta \Omega \Delta E N_{t} \varepsilon^{syst}_{i} },
\label{Igdhr3}
\end{equation}
where $\Delta\Omega$ is the solid angle from the target to the neutron detector, $\Delta E$, is the width of the neutron energy bin, $N_{t}$ is the $^3$He target thickness determined to be (8.3$\pm$0.3)$\times$10$^{21}$ atoms/cm$^{2}$ and $\varepsilon^{syst}_{i}$ is the system efficiency accounting for both the intrinsic efficiency of the neutron detector and the neutron multiple scattering effect calculated at the $i^{th}$ energy bin using a GEANT4~\cite{geant} simulation of the experiment.

Two types of systematic uncertainties were identified: the bin-dependent and the overall normalization uncertainties. The former were asymmetric and arose from the PH cuts on the neutron spectra. The latter were bin-independent, symmetric and the major contributors from most to least important were: $\delta$P$_b$ (5\%), $\delta$P$_t$ (4.2\%), $\delta$N$_\gamma$ (4.2\%) (for which the main contribution was from the deuteron photodisintegration cross section uncertainty (3.0\%)~\cite{d2o}), $\delta$N$_t$ (4.0\%), $\delta\varepsilon^{syst}_{i}$ (2.8\%)~\cite{Trotter, setze} and $\delta\Delta\Omega$ (2\%).

The spin-dependent double differential cross sections for the extended target obtained at an incident photon energy of 16.5 MeV for both spin-helicity states as a function of neutron energy (E$_n$) are shown in Fig.~\ref{fig:165}. The solid and dashed curves are the GEANT4 simulation results using the calculations based on Ref.~\cite{Deltuva} and Ref.~\cite{Skibinski}, respectively. The band in each panel shows the overall systematic uncertainties combined in quadrature. Although the magnitudes of the double-differential cross sections are overall larger in the parallel than those in the antiparallel spin-state, the distributions are not well described by either of these calculations. An excess of neutron events was observed close to the end-point energies at laboratory scattering angles of $30\,^{\circ}$, $45\,^{\circ}$, $150\,^{\circ}$ and $165\,^{\circ}$ due to large backgrounds. These energy bins were removed and their contribution to the overall strength of the distributions which was found to be $\sim$ 1\% for both spin-states and all scattering angles, was added heuristically based on the theory.

\begin{figure}[!h]
  \centering
    \includegraphics[width=0.5\textwidth]{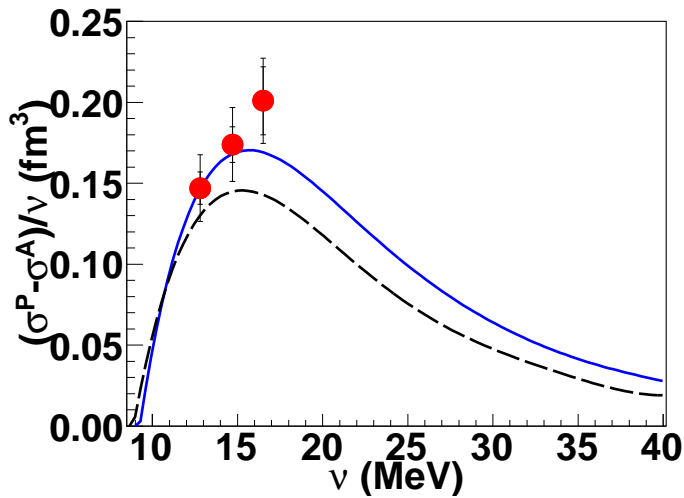}
    \caption{(Color online) The GDH integrand results compared with the theoretical predictions of Ref.~\cite{Deltuva} (solid-blue curve) and Ref.~\cite{Skibinski} (dashed-black curve). The inner error bars of the data points represent the statistical uncertainties while the outer include both the statistical and systematic uncertainties added in quadruture.}
   \label{fig:gdh}
\end{figure}

Iterative Monte Carlo simulations using GEANT4 were carried out in order to correct the spin-dependent double differential cross section distributions for finite-geometry effects. The resulting distributions were integrated over the neutron energy to extract the single differential cross sections. The unmeasured part of the distributions for E$_n<$2 MeV was added based on the theoretical distributions including the Coulomb interaction which were normalized to the magnitude of the first valid neutron bin (2.0-2.5 MeV) for both states and all angles. Legendre polynomials up to the 4$^{th}$ order were used to fit the single differential cross sections distributions for both states. To achieve the fit with the highest statistical signifigance, the single differential cross section points corresponding to the angle of $105\,^{\circ}$ were removed. The $\chi^2/$(degrees of freedom) for the fit at the parallel (anti-parallel) state was found to be 1.01 (1.39). The fitting curves were integrated over the angle to extract the spin-dependent total cross sections and the value of the GDH integrand. More details about this analysis can be found in Refs.~\cite{GeorgePRL,laskaristhesis}.

Table~\ref{table:gdh} summarizes the spin-dependent total cross sections and the contribution from the three--body photodisintegration to the $^3$He GDH integrand together with the predictions based on the models presented in Ref.~\cite{Deltuva} and Ref.~\cite{Skibinski}. Differences between the measured spin-dependent total cross sections and the calculated values are found at the incident photon energy of 16.5 MeV. This is in contrast to a very good agreement observed between the previous measurements~\cite{GeorgePRL,laskaristhesis} and the calculations based on Ref.~\cite{Deltuva} at 12.8 and 14.7 MeV. The measured GDH integrand at 16.5 MeV was found to be slightly more than one standard deviation larger than the maximum calculated value based on Ref.~\cite{Deltuva}. Fig.~\ref{fig:gdh} shows the contributions of the three--body photodisintegration of $^3$He to the GDH integrand together with the theoretical predictions based on Refs.~\cite{Deltuva,Skibinski} as a function of the incident photon energy. To investigate whether the larger than expected GDH integrand value at 16.5 MeV is due to statistics, future measurements at higher energies are needed. The outcome of these measurements will have important implications for the GDH integral of $^3$He below the pion threshold.
\begin{table}[!ht]
\begin{center}
\caption{Total cross sections, $\sigma^{P}$ and $\sigma^{A}$ and the GDH integrand, $(\sigma^{P}-\sigma^{A})/\nu$, with statistical uncertainties followed by systematics, compared with theoretical predictions.}
\begin{tabular*}{0.48\textwidth}{@{\extracolsep{\fill} }c c c c }
\hline
\hline
\backslashbox{ }{ } & $\sigma^{P}$($\mu$b) & $\sigma^{A}$($\mu$b) & $(\sigma^{P}-\sigma^{A})/\nu$ (fm$^{3}$) \\ \hline
\hline
Experiment & 933(12)(100) & 764(12)(91) & 0.201(0.021)(0.016) \\ 
Ref.~\cite{Deltuva} & 1077 & 935 & 0.169 \\ 
Ref.~\cite{Skibinski} & 1099 & 979 & 0.143 \\ \hline
\end{tabular*}
\label{table:gdh}
\end{center}
\end{table}

\begin{figure}[!h]
  \centering
    \includegraphics[width=0.5\textwidth]{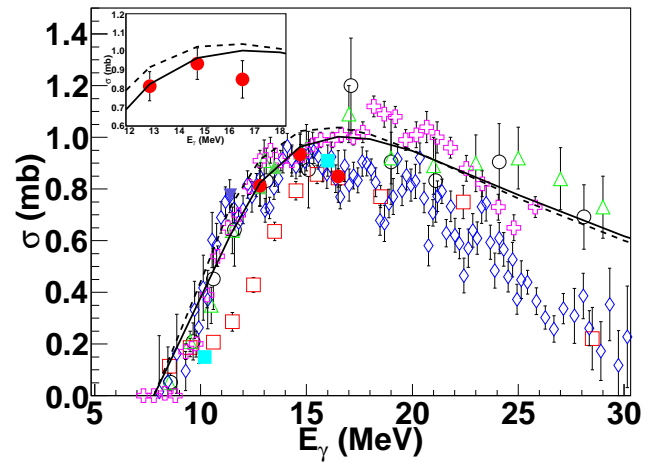}
    \caption{(Color online) All currently available total cross section data for the ${^3}He(\gamma,n)pp$ reaction up to 30 MeV: Refs.~\cite{GeorgePRL} and data presented for first time in this letter (filled circles), Ref.~\cite{gorbunov} (open circles), Ref.~\cite{gerstenberg} (open squares), Ref.~\cite{berman} (diamonds), Ref.~\cite{gorbunovproc} (open upward triangles), Ref.~\cite{faul} (open crosses), Ref.~\cite{naito} (filled squares), Ref.~\cite{perdue} (filled upward triangles), Ref.~\cite{zong} (filled donward triangle) in comparison to the calculations from Refs.~\cite{Deltuva} (solid curve) and Refs.~\cite{Skibinski} (dashed curve). In the insert, the data by our collaboration are shown and compared with the theories. The older measurements~\cite{gorbunov,gerstenberg,berman,faul} are presented with the statistical uncertainties while the newer data points~\cite{GeorgePRL,naito,perdue,zong} include both the statistical and systematic errors added in quadruture.}
   \label{fig:txs}
\end{figure}

The unpolarized cross section was extracted as the average of the spin-dependent cross sections and was found to be equal to 849$\pm$9$\pm$100 $\mu$b. Fig.~\ref{fig:txs} shows all unpolarized total cross sections data up to 30 MeV compared to the total cross section calculations from Ref.~\cite{Deltuva} (solid curve) and Ref.~\cite{Skibinski} (dashed curve). A general agreement between the two calculations and the experimental data can be observed for incident photon energy below 15 MeV. A serious discrepancy can be seen between different sets of data above 15 MeV while our result agrees with the measurements of Refs.~\cite{berman,gerstenberg} and the most recent data of Ref.~\cite{naito} which favor smaller total cross section above 15 MeV. In order to resolve this discrepancy and to further quantify the three--body contribution to the GDH integral, measurements above 16.5 MeV for this channel are necessary. These measurements combined with the recently acquired data from the two-body photodisintegration channel~\cite{laskaristhesis} will constrain the contribution to the GDH integral for $^3$He below the pion threshold. 

This work is supported by the U.S. DOE under contract numbers DE-FG02-03ER41-231,-033,-041, Duke University and the PNSC under Grant DEC-2013/10/M/ST2/00420. The numerical calculations of Krak\'ow group were performed on the clusters of the JSC.

\end{document}